\documentclass[showpacs,prb,twocolumn,superscriptaddress]{revtex4}
\usepackage{epsfig}
\usepackage{bm}

\begin{document}

\title{Valence-Bond Crystal, and Lattice Distortions in a Pyrochlore Antiferromagnet with Orbital Degeneracy.}
\author{S. Di Matteo}
\affiliation{Laboratori Nazionali di Frascati INFN, via E. Fermi 40, C.P. 13, 
I-00044 Frascati (Roma) Italy}
\affiliation{Dipartimento di Fisica, Universit\`a di Roma III, via della Vasca
  Navale 84, I-00146 Roma Italy}
\author{G. Jackeli}
\altaffiliation[Also at ] {E. Andronikashvili Institute of Physics,
Georgian Academy of Sciences, Tbilisi, Georgia.} 
\affiliation{Ecole Polytechnique F\'ed\'erale de Lausanne,
Institute for Theoretical Physics, CH-1025, Lausanne,
Switzerland}
\affiliation{Institut Laue Langevin, B. P. 156, F-38042,
 Grenoble, France}
\author{N. B. Perkins}
\affiliation{Laboratori Nazionali di Frascati INFN, via E. Fermi 40, C.P. 13, 
I-00044 Frascati (Roma) Italy}
\affiliation{Bogoliubov Laboratory of Theoretical Physics, JINR, 141980, Dubna, Russia}
\date{\today}
\begin{abstract}
We discuss the ground state properties of a spin 1/2 magnetic ion with 
threefold $t_{2g}$ orbital degeneracy on a highly frustrated pyrochlore lattice, 
like Ti$^{3+}$ ion in B-spinel  MgTi$_2$O$_4$. We formulate an effective  
spin-orbital Hamiltonian and study its low energy sector by constructing several exact-eigenstates in the limit of vanishing Hund's coupling. We find that orbital degrees of freedom modulate 
the spin-exchange energies, release the infinite spin-degeneracy of pyrochlore structure,
and drive the system to a non-magnetic spin-singlet manifold. The latter is a collection of 
spin-singlet dimers and is, however,  highly degenerate with respect of dimer orientations.
This ``orientational'' degeneracy  is then lifted by a
magneto-elastic interaction that optimizes the previous energy gain 
by distorting the bonds in suitable directions and leading to a tetragonal phase. 
In this way a valence bond crystal state is formed, through the condensation of dimers along helical chains running around the tetragonal $c$-axis, as actually observed in MgTi$_2$O$_4$.
The orbitally ordered  pattern in the dimerized phase is predicted to be  
of ferro-type along the helices and of  antiferro-type  between them. 
Finally, through analytical considerations 
as well as numerical ab-initio simulations, we predict a possible experimental 
tool for the observation of such an orbital ordering, through resonant x-ray scattering.
\end{abstract}
\pacs{75.10.Jm, 75.30.Et }
\maketitle

\section{Introduction.}

A frustrated antiferromagnet is characterized by the topology of underlying lattice and/or 
by the presence of competing interactions 
that preclude every pairwise magnetic interaction to be satisfied at the 
same time. Such physical systems have recently
attracted a wide interest,\cite{1} due to the concept 
of "macroscopic" degeneracy, namely the existence of a huge number of  states with the same energy.
This degeneracy in the  ground state 
manifold can be usually removed through a large 
variety of effects, like order-out-of-disorder mechanisms \cite{villain}
by thermal or quantum fluctuations, or spin-Peierls like
symmetry-lowering transitions.\cite{ueda,tcher}
In some highly frustrated models, like nearest-neighbor 
Heisenberg antiferromagnet on pyrochlore lattice,
the former mechanisms are inactive\cite{moessner}  and 
such a spin system would remain liquid
down to the lowest temperatures,\cite{sl}
unless magnetoelastic couplings induce a symmetry-breaking transition.
Yet, in real compounds  geometrical frustration can also be partially or 
fully released when magnetic ions forming a frustrated lattice possess an orbital degeneracy. 
Typically this happens in transition metal ions with orbitally degenerate partly-filled $d$-levels.  
The physical behavior of such systems is expected to 
 be drastically different from that of pure spin models, as the occurrence
of an orbital ordering (OO) can modulate the spin exchange and  lift the 
geometrical degeneracy of the underlying lattice. 
Indeed, a variety of novel phases driven by orbital degrees of freedom 
can be stabilized in this way.
Among known  examples are  vanadium $d^2$ compounds with frustrated lattices, 
where the orbital order is shown to induce a spin-singlet
ground state without any long-range magnetic order for triangular
lattice,\cite{pen} or a spin ordered one, for pyrochlore lattice.\cite{TS} 

In this paper we study a system with  threefold-orbitally-degenerate $S=1/2$ 
magnetic ions in a corner-sharing tetrahedral (pyrochlore) lattice. Our work is motivated by the
very recent synthesis\cite{isobe} and interesting 
experimental data on $B$-spinel MgTi$_2$O$_4$,\cite{isobe,schmidt} 
a $d^1$-type transition metal oxide. Here  magnetically
active Ti$^{3+}$ ions form a pyrochlore lattice and are characterized by a
single electron in a $d$-shell. The crystal field of the oxygen octahedron 
surrounding each Ti ion splits this $d$-level  into a high energy doublet 
$e_{g}$ and a low energy triplet $t_{2g}$ in which $d$-electron resides.\cite{trig}
The ground state of  a spin one-half  Ti$^{3+}$ ion is thus
threefold-orbitally-degenerate.
The recent experiments on  MgTi$_2$O$_4$ have  shown
that this compound
 undergoes a metal-to-insulator 
transition on cooling below 260 K, with an associated 
cubic-to-tetragonal lowering of the symmetry.\cite{isobe} 
At the transition the magnetic susceptibility continuously  decreases and saturates, in the
insulating phase, to a value which is anomalously small for  spin $1/2$ local moments:
for this reason the insulating phase has been interpreted as a spin-singlet phase. 
Subsequent synchrotron and neutron powder diffraction
experiments have revealed that the low-temperature crystal
structure is made of alternating short and long Ti-Ti bonds
forming a helix about the tetragonal $c$-axis.\cite{schmidt}
These findings have suggested a removal of the pyrochlore degeneracy
by a one-dimensional (1D) helical dimerization of the spin pattern, 
with spin-singlets (dimers) located at short bonds. This phase can be regarded  as a 
valence bond crystal (VBC) since the long-range order of spin-singlets 
extends throughout the whole pyrochlore lattice.

The aim of the present work is to discuss the microscopic mechanism
behind the realization of this unusual and intriguing VBC structure
on the pyrochlore lattice.  We argue that the key role in this mechanism is
played by orbital degeneracy and, remarkably, one does not have to invoke any 
additional exotic interaction to stabilize such a novel phase, as necessary for purely spin models.
Indeed, due to the orbital degeneracy, various type of 
spin-singlet phases, such as resonating valence bond  (RVB)  and VBC states 
can be formed even for unfrustrated cubic lattice.\cite{feiner}

The orbital degree of freedom does modulate spin exchange
energies,  thus removing the infinite spin degeneracy, characteristic 
of pyrochlore structures, and drives the system to a non-magnetic
spin-singlet state. This latter is a collections of spin-singlet dimers
with a residual macroscopic degeneracy of orientational character. 
The residual degeneracy is then lifted by a magnetoelastic interaction,
that optimizes superexchange energy gain by distorting each tetrahedron
in such a way as to lead to the experimentally observed helical
pattern in MgTi$_2$O$_4$. We also find that the  helical dimerized state is 
accompanied by a peculiar orbital pattern in which orbital order is of 
ferro-type along the helices and of  antiferro-type between them. 

Moreover, we can show how to identify such an orbital ordering  experimentally, 
by means of resonant x-ray scattering (RXS). In fact, differently from what happens in the case of 
manganites,\cite{elfimov,natoli,igarashi,tapan} where the ratio between 
the OO-induced and Jahn-Teller (JT) induced effects is about 1/10 in amplitude, 
for MgTi$_2$O$_4$ this ratio is about 1/3, due to the less 
distorted oxygen octahedra, as well as to the presence of $t_{2g}$ electrons instead of $e_g$, 
that couple less to the oxygen environment. Such a reduced ratio allows a subtle interference 
effect between OO and JT terms, 
that gives rise to an increase of the signal by a factor of about 1.7.
We have performed a detailed analytical analysis as well as an 
ab-initio numerical simulation
to suggest some experiments in this direction.

In more detail, the paper is organized as follows: in Section II we derive
an effective Kugel-Khomskii\cite{kugel} model Hamiltonian, 
and deduce its  possible low-energy states on the pyrochlore lattice. 
We then introduce the spin-singlet ground state
manifold and discuss its degeneracy due to the dimer coverings of the
pyrochlore lattice. 
 In Section III we  consider the magnetoelastic interactions of the pyrochlore lattice
and analyze the effects of the coupling of bond distortions with the orbitally driven
exchange modulation, 
first qualitatively, and then quantitatively, underlining the influence 
of the spin-singlet  phase on the strength of such a distortion. 
In Section IV the effect of an applied magnetic field is briefly discussed, and, finally, in Section V we analyze the 
crystal structure and the orbital symmetry of the tetragonal phase, deriving
the analytic expression for the measurable quantities in a RXS experiment. 
We also perform a series of numerical 
{\it ab-initio} simulations by means of the finite difference method, 
implemented in the FDMNES package,\cite{yvesprb} 
in order to propose a possible experiment to detect the orbital pattern. 
To faciliate the reader, two Appendices, A and B, are given 
for technical details.

Part of the results presented here were already announced in a previous
short communication.\cite{prl}

\section{Model and formalism.}

\subsection{Effective spin-orbital Hamiltonian.}

 Here we  discuss the superexchange spin-orbital  Hamiltonian for threefold
orbitally-degenerate $d^1$-ions on a pyrochlore lattice.
We assume that the insulating phase of MgTi$_2$O$_4$ is of Mott-Hubbard type and can, thus, be 
described by the Kugel-Khomskii model.\cite{kugel}
We consider the system in its cubic structure and look for possible instabilities 
towards symmetry reductions.

The relevant electronic degrees of freedom
are described by spin $S=1/2$ and pseudospin $\tau=1$ operators.
This latter labels the orbital occupancies of $t_{2g}$ orbitals, 
($|\alpha\beta\rangle=|xy\rangle,~|xz\rangle,~|yz\rangle$), 
with  the correspondence: $\tau^{z}=-1\rightarrow|yz\rangle$, $\tau
^z=0\rightarrow |xy\rangle$,  and $\tau ^z=1\rightarrow |xz\rangle$.  
Our parameters are the nearest-neighbor (NN) electron hopping  matrix $\hat{t}$, defined
 in the Appendix A Eq. (\ref{hopping}),  
the Coulomb on-site repulsions $U_1$ (within the same 
orbital) and $U_2$ (among different orbitals), and the Hund's exchange, $J_H$.
For $t_{2g}$ wavefunctions the relation  $U_1=U_2+2J_H$ holds due to rotational
symmetry in real space. 

Considering the pyrochlore structure the effective
Hamiltonian can be simplified by retaining only
the leading  hopping parameter, $t$, due to the NN
$dd\sigma$ overlap and neglecting the smaller $dd\pi$ and $dd\delta$ contributions. This is justified by the fact that the transfer integrals due to the 
$\pi$ and $\delta$ bonding are, respectively, around $1/10$ and $1/3$ of that of $\sigma$ 
bonding.\cite{t} 
The major advantage of this simplification, 
is that $dd\sigma$ overlap  in $\alpha\beta$ plane 
connects only  the corresponding orbitals of the same $\alpha\beta$ type.
Thus, the total number of electrons in each orbital state is 
a conserved quantity  and, therefore, the orbital part of the effective 
Hamiltonian (\ref{spinorb2}) becomes Ising-like (or better, a Potts Z$_3$-like), 
making possible to get analytical results about the ground-state. 
The spin-orbital Hamiltonian is presented in Appendix A in the most general
form [see Eq.(\ref{spinorb2})], when also $dd\pi$ and $dd\delta$ hopping elements are considered.
Here we introduce the simplified version with only $dd\sigma$ hopping terms:
\begin{eqnarray}
&&H_{\rm eff}=  
-J_1\sum_{\langle ij\rangle} {\big [}\vec S_i\cdot \vec S_{j} +3/4 {\big
  ]}O_{ij}
\label{spinorb}\\
&&+J_2\sum_{\langle ij\rangle} {\big [}\vec S_i\cdot \vec S_{j} -1/4 {\big ]}O_{ij}+
J_3\sum_{\langle ij\rangle} {\big [}\vec S_i\cdot \vec S_{j}-1/4 {\big ]}\tilde{O}_{ij}
\nonumber
\end{eqnarray}
where the sum is restricted to the NN  sites on the pyrochlore lattice.
The orbital contributions along the bond $ij$ in $\alpha\beta$-plane is 
given  by 
\begin{eqnarray}
O_{ij}&=&P_{i,\alpha\beta}(1-P_{j,\alpha\beta})+P_{j,\alpha\beta}(1-P_{i,\alpha\beta})\nonumber\\
\tilde{O}_{ij}&=&P_{i,\alpha\beta}P_{j,\alpha\beta}, 
\label{orbpart}
\end{eqnarray}
where $P_{i,\alpha\beta}=T_{i}^{aa}$ stands for the projector
on orbital state $|a\rangle=|\alpha\beta\rangle$ and is defined in Appendix A.

The first and second terms in $H_{\rm eff}$ (\ref{spinorb}) describe the 
ferromagnetic (FM)  $J_1=t^2/(U_2-J_H)$ and 
the antiferromagnetic (AFM) $J_2=t^2/(U_2+J_H)$ interactions,
respectively, and are active only when the two sites involved are occupied by 
different orbitals. The last term is AFM, with
$J_3=\frac{4}{3}t^2{\big [}2/(U_2+J_H)+1/(U_2+4J_H){\big ]}$, and 
is non-zero only when the two sites have the same orbital occupancy.
Parameters that play a role in the Hamiltonian (\ref{spinorb}) are: 
 $t\equiv t_{\sigma} \simeq
0.32$ eV,
$J_H \simeq 0.64$ eV and $U_2\simeq 4.1$ eV.\cite{t,mizo}
Thus $\eta =J_H/U_2\simeq
0.15\ll 1$ and, just in order to present the
results in a more transparent form, we expand the exchange energies  around $\eta=0$.
We get $J_1\simeq J(1+\eta ) $, $J_2\simeq J(1-\eta )$ and $J_3\simeq
4J(1-2\eta )$ where $J=t^2/U_2 \simeq 25$ meV represents the overall energy
scale. 

The realization of the spin-orbital model (\ref{spinorb}) on the 
pyrochlore lattice has an immediate and important consequence, 
namely, that only some bonds can contribute to the energy, depending on their orbital configuration. Indeed, every bond $ij$ in the $\alpha\beta$ 
plane has zero energy gain unless at least one of 
the two sites $i$ and $j$ has an occupied orbital of $\alpha\beta$ kind. 
The strength, as well as the sign, of spin-exchange energy associated with
two NN sites $i$ and $j$ depends only on their orbital occupations and the direction of the $ij$ bond. 
We can thus classify tetrahedral bonds in four types: 

(i) $b_0$, when both ions at sites $i$ and $j$ of the generic $\alpha\beta$-plane have $\alpha\beta$ orbital occupancy (Fig. 1a). 
It is characterized by a Hamiltonian with strong AFM exchange $\sim J$: 
\begin{equation}
H_{b_0}=-J(1-2\eta)(1-4 \vec S_i\cdot \vec S_{j}).
\label{b0}
\end{equation}

(ii) $b_1$, if the two sites of bond $ij$ in $\alpha\beta$-plane are occupied by one $\alpha\gamma$ and one $\alpha\beta$ orbitals,
 $\gamma\not=\beta$(Fig. 1b). This bond has a weak FM exchange $\sim \eta J$ and the
corresponding Hamiltonian is:
\begin{equation}
H_{b_1}=-J(1+\eta/2+2\eta \vec S_i\cdot \vec S_{j}).
\label{b1}
\end{equation}

(iii) $b_2$, with the two sites of bond $ij$ in $\alpha\beta$-plane occupied by one $\alpha\gamma$ and one $\beta\gamma$ orbitals. 
In this case there is no energy contribution: $H_{b_2}=0$.

(iv) $b_3$, if both sites of bond $ij$ in $\alpha\beta$-plane are occupied by two $\alpha\gamma$ ($\beta\gamma$) orbitals. 
Also this bond does not contribute to the energy: $H_{b_3}=0$.

\begin{figure}
\epsfysize=14mm
\centerline{\epsffile{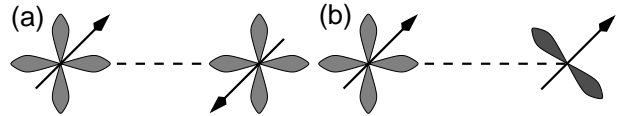}}
\caption{Orbital  arrangements on a bond: 
(a) bond $b_0$ with strong AFM coupling $\sim J$, 
(b) bond $b_1$ with weak FM coupling $\sim \eta J$.}
\label{fig1}
\end{figure}

Thus, only  $b_0$- and  $b_1$-bonds have a binding energy 
and for any given orbital configuration on the pyrochlore lattice the
Hamiltonian (\ref{spinorb}) is just a linear combination of $H_{b_0}$ and $H_{b_1}$. 
Moreover, the explicit form of Eqs. (\ref{b0}) and (\ref{b1}) 
implies that, in the limit $\eta\rightarrow 0$, a set of exact eigenstates of (\ref{spinorb}) can 
be constructed: in fact, orbital interactions are already diagonal, and,
in this limit, the only Heisenberg term that survives is that of $H_{b_0}$. This latter, by definition, is active only along the direction connecting the two sites with equally filled $\alpha\beta$ orbitals. 
This implies that the only "infinite" configuration of interacting spins is obtained when the spin-interactions 
act along the same direction, i.e.,
the AFM Heisenberg chain, which is exactly soluble. \cite{1D} 

In the next subsection, starting from an isolated tetrahedron we discuss possible orbital arrangements 
on the pyrochlore lattice and provide the solution of corresponding spin Hamiltonian.

\subsection{Possible phases on a pyrochlore lattice and their energetics}

\begin{figure}
\epsfysize=47mm
\centerline{\epsffile{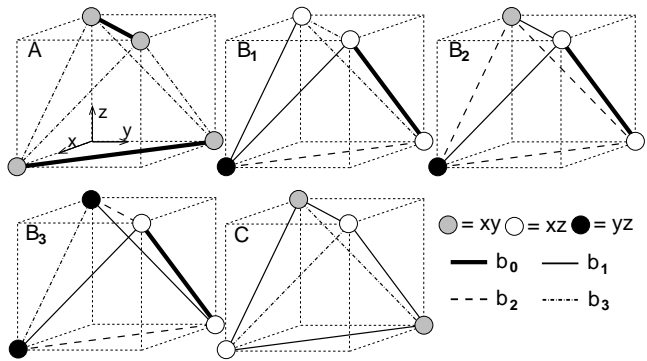}}
\caption{Orbital and bond arrangements on a tetrahedron for cases $A$,
  $B_{1,2,3}$ and $C$ discussed in the text.}
\label{fig2}
\end{figure}

It is possible to show that only 
three energetically inequivalent tetrahedrons can 
be singled out of the four $b_{n}$ bonds introduced above.\cite{prl}
Topologically, they can be characterized by the number 
of strongest $b_0$-bonds and classified in three 
following types [pictorially shown in Fig. \ref{fig2}]: 
$A$-type tetrahedra, with two $b_0$ bonds, 
$B$-type tetrahedra with one $b_0$ bond and $C$ tetrahedra, with no $b_0$-bonds. 
Notice that, whatever the orbital configuration is, every single tetrahedron must obey the constraint $2n_{b_0}+n_{b_1}=4$. 
This implies that $A$ tetrahedra have no $b_1$-bonds, 
$B$-type tetrahedra are characterized by 2 $b_1$ bonds and, for $C$ tetrahedra, 
there must be 4 $b_1$-bonds.
We can further classify $B$-type 
tetrahedra, according to the choice of the two orbitals 
on the tetrahedron bond opposite to the $b_{0}$-bond, into:
 $B_1$, $B_2$ and $B_3$ (see Fig. \ref{fig2}) where, respectively, the bond opposite 
to $b_0$ one is of $b_1$, $b_2$ and $b_3$ type.
We stress again that the number of $b_2$ and $b_3$ bonds is not relevant as far as the Hamiltonian (\ref{spinorb}) 
is considered: 
they acquire importance only when magnetoelastic correlations among bonds will be introduced, in the next section,
 as they allow to distinguish topologically the various configurations in the pyrochlore lattice.
$A$, $B$, and $C$ tetrahedra are the bricks that allow to build the orbital pattern throughout
the whole pyrochlore lattice. Because of the Ising-form of orbital
interactions, in the following we can focus simply on these three cases, 
relying on the fact that configurations with a linear superposition of
orbitals on each site  must have a higher energy. 
We shall do only one exception to study a case with a
particular physical meaning, i.e., that of a "cubic" symmetry, where each site
is occupied by a linear superposition with equal weights 
of the three orbitals $\frac{1}{\sqrt{3}}[|xy\rangle+|xz\rangle+|yz\rangle]$.

If we now try to cover the whole pyrochlore lattice with these tetrahedra, 
we can have the following global patterns:

(I) {\bf Heisenberg chains}: If all tetrahedra of pyrochlore lattice are characterized by two $b_0$ bonds (i.e., all sites are occupied 
by the same orbital), then the effective Hamiltonian
(\ref{spinorb}) can be mapped into a set of one dimensional decoupled AFM 
Heisenberg chains that lie within the plane of the chosen orbital. 
The only interactions are due to $b_0$-bonds and are described by the spin Hamiltonian 
(\ref{b0}).
Thus, the ground-state energy per site can be evaluated exactly
 by using the results for an Heisenberg chain,\cite{1D} that give $E_{A}=-2.77 (1-2\eta)J$.

\begin{figure}
\epsfysize=130mm
\centerline{\epsffile{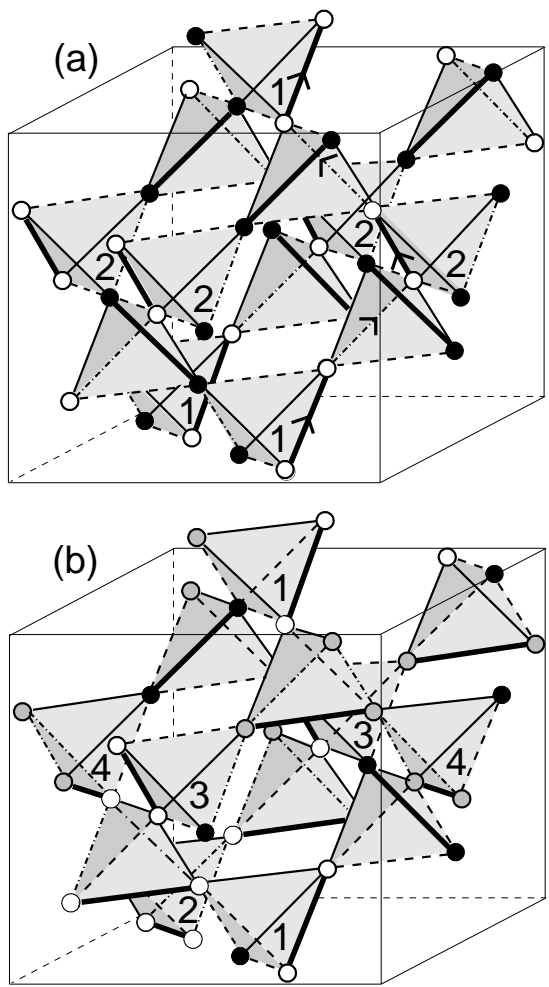}}
\vspace{0.5cm}
\caption{The ground state coverings of the unit cubic cell through dimers. 
Locations of singlets are represented by thick links.
Different numbers correspond to inequivalent tetrahedra.
(a) The helical dimerization pattern (indicated by arrows)
is formed by alternating short $b_0$ and long  $b_3$ bonds,
 Dimer phase  $B_3$.
(b) One of the possible coverings of the cubic unit cell 
by $B_1B_2$ tetrahedra.}
\label{vbc}
\end{figure}

(II) {\bf Dimer phase $B$}: This state is made of only $B$-type tetrahedra
 with one strong $b_0$-bond, and two intermediate $b_1$-bonds. 
As all three $B_i$ tetrahedra are energetically equivalent 
all possible coverings of pyrochlore lattice by $B_i$ tetrahedra have the
same energy. When pyrochlore lattice is covered by   $B_i$-types
tetrahedra (two possible  coverings  are shown in Fig. \ref{vbc}),
then  each spin is engaged in one strong AFM $b_0$ bond and two weak FM
$b_1$-bonds. Such coverings form a degenerate manifold and the corresponding energy can
be calculated as follows. 
In the limit $\eta \rightarrow 0$, the spin-only 
Hamiltonian can be solved exactly, as it can be decomposed into a sum 
of spin-uncoupled $b_0$ bonds. In this case the energy minimum is 
reached when the Heisenberg term of the $b_0$-bond is the lowest, i.e., 
for a pure quantum spin-singlet ($\vec S_i\cdot \vec S_{j}=-3/4$).
Remarkably, such spin-singlet (dimer) states, in the limit $\eta \rightarrow 0$, 
are also exact eigenstates of the full 
Hamiltonian (\ref{spinorb}).
As $\eta \ll 1$, the dimer state is stable against  
the weak FM interdimer interaction. In this case the magnetic  
 contribution along the FM $b_1$-bond is zero ($\langle \vec{S}_i\cdot \vec{S}_{j}\rangle=0$
for $i$ and $j$ belonging to different dimers)
  and we are led to an energy 
per site given by: 
$E_B=E_{b_0}/2+E_{b_1}=-(3-\frac{7}{2}\eta)J$. Here $E_{b_{0(1)}}$ is the energy
of the bond $b_{0(1)}$.

(III) {\bf FM order}: In this case all tetrahedra are of $C$-type
with four interacting $b_1$ bonds and two $b_3$ (or one $b_2$ and one $b_3$) 
noninteracting bonds. Thus, all non zero spin-exchanges are ferromagnetic and given by Eq. (\ref{b1}).
The ground state for this type of orbital ordering is thus ferromagnetic and 
has an energy per site given by $E_C=2E_{b_1}=-2(1+\eta)J$.

(IV) {\bf Frustrated AFM}: Even if the "Ising" form of orbitals interaction 
implies that configurations with linear superposition 
of orbitals on each site must have a higher energy, we consider for completeness the case where each orbital is occupied 
by a linear superposition with equal weight of the three orbitals: $\frac{1}{\sqrt{3}}(|xy\rangle+|xz\rangle+|yz\rangle)$.
The realization of this phase restores the full pyrochlore lattice symmetry 
and, after averaging Eq. (\ref{spinorb}) over the orbital configurations on neighboring sites $i$ and $j$, the system is described 
by the spin Heisenberg Hamiltonian $H_D/J=\sum_{ij}[-5/9+(4/9-16\eta/9) \vec{S}_i\cdot \vec{S}_j]$, whose ground-state energy 
per site is $E_{D} \simeq -(1.89 -0.89 \eta)J$. Here we borrowed the numerical evaluation for ground-state energy in a pyrochlore lattice 
from Ref. [\onlinecite{canals}]: $(1/N)\sum_{ij} \vec{S}_i\cdot \vec{S}_j\simeq -0.5$. Such a state is highly frustrated and its ground state is a spin liquid.\cite{sl}
It seems worthwhile to note that the energy in this phase is higher than that of the other phases, as it does not exploit at all the potential 
energy-gain contained in the orbital ordering.\cite{spinorbit}

\begin{figure}
\epsfysize=65mm
\centerline{\epsffile{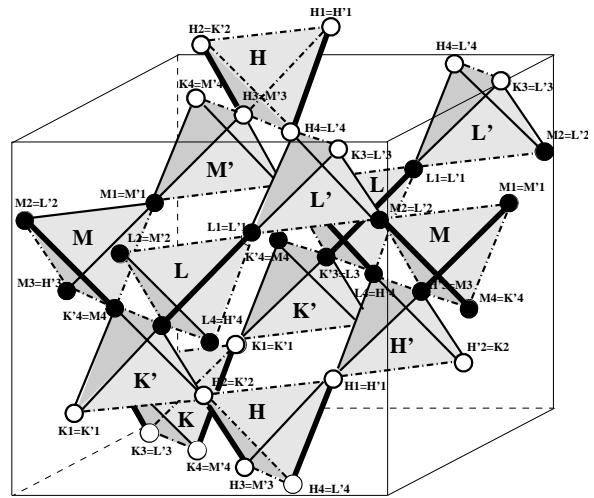}}
\vspace{0.5cm}
\caption{The coverings of the unit cubic cell through the tetrahedra of 
A-type with two singlets and C-type with no singlets. Tetrahedra are labeled as in Fig. \ref{plconstr}.}
\label{mixed}
\end{figure}

(V) {\bf Mixed (dimer) phase $AC$}: 
It is of course possible to cover the pyrochlore 
lattice also by means of mixed configurations of tetrahedra. 
There are in principle infinite possibilities in this sense, 
but those minimizing the energy must contain, in average, 
at least one singlet bond every tetrahedron, 
due to the big energy gain related to the spin-singlet state in H$_{b_0}$.
While it is not possible to have, in average, more than one spin-singlet bond every tetrahedron,\cite{nota} one can fill the whole pyrochlore lattice 
by means of a mixed configuration  made of alternated tetrahedra with 
two and zero $b_0$-bonds ($A$- and $C$- type tetrahedra, respectively).
This configuration is shown in Fig.\ref{mixed}. 
For this type of orbital arrangement one can also construct an exact ground state 
of corresponding spin 
Hamiltonian, in the limit of vanishing Hund's coupling. The ground state is given by dimer phase where
there are two spin-singlets, located opposite each other, on strong $b_0$ bonds 
of $A$-type tetrahedra while $C$-type tetrahedra has no singlets.  
As in average,  $n_{b_0}=1$ and $n_{b_1}=2$, this configuration is degenerate with Dimer phase $B$, 
as far as only $dd\sigma$ overlap is considered.

\subsection{Ground State Manifold}

Energies of the possible phases on the pyrochlore lattice 
can be discussed in terms of $\eta$, the only free parameter available.
For $\eta =0$ the lowest ground-state energy is that of spin-singlet 
degenerate manifold.
With increasing $\eta$ we find only one phase transition
 at $\eta_c=2/11\simeq 0.18$, from spin-singlet manifold 
to a FM phase. As $\eta_c$ is above our estimated value of 
$\eta\simeq 0.15$, we can conclude that the ground state manifold   
is given by spin-singlet phases spanned by  degenerate dimer states of $B$ or $AC$-types.
This degenerate manifold is characterized by a static pattern of spin-singlets (dimers)
throughout  the whole pyrochlore lattice and, thus, is different from RVB state.
Each dimer covering  is frozen in an  exact eigenstate of the
Hamiltonian (\ref{spinorb}) for $\eta=0$. For finite $\eta$ the different dimer
patterns are not connected by the Hamiltonian:  the bond corresponding to the dimer
in each tetrahedron is fixed, being determined by orbital pattern
and orbital degrees of freedom are static  variables. 
Thus a  tunneling between different dimer states can not take place.

In this  spin-singlet manifold the original spin degeneracy is removed.
However there is still a remaining degeneracy  to be lifted. 
The degeneracy  of $B$-manifold   
is related to the freedom in the choice of the two orbitals on the 
tetrahedron bond opposite to the one of the singlet. 
Different choices of these orbitals give rise to inequivalent dimer covering
patterns of the pyrochlore 
lattice with one dimer per tetrahedron  (see Fig. \ref{vbc}).
This degeneracy is given by the number of such dimer coverings and the
corresponding number of states can be estimated as follows.
In the $B$-manifold there is one singlet per tetrahedron and each spin can be engaged in only
one singlet. Moreover  when a singlet is located on a given tetrahedron 
then each neighboring tetrahedron is left with only three possible choices for a
  singlet location. Thus the number of such coverings grows with the
system size as ${\cal N}_{B} \sim 3^{N_T}=\sqrt{3}^{N}$. 
Here $N_T=N/2$ is the number of tetrahedra 
and we have ignored the contributions coming from closed loops (hexagons) on the pyrochlore
lattice. 

The asymptotical determination of  $AC$-manifold degeneracy is a more difficult task. 
However, based on the simple arguments we can 
estimate its lower and upper limit correctly. 
First, we divide the pyrochlore structure in two sublattices formed by two differently oriented
tetrahedra: one sublattice is composed by $A$-type tetrahedron with two  singlets located 
opposite each other  and the other by $C$-type tetrahedron with no singlets.
Then consider an $A$ tetrahedron with two singlets in $\alpha\beta$ plane: it can be directly verified that the four NN tetrahedra of $A$ sublattice, connected to it by straight lines in $\alpha\beta$ plane, will have only two possible orbital choices. The remaining eight NN tetrahedra of the $A$ sublattice have, instead, either two or three choices to locate the two spin-singlets. These constraints are dictated by the condition that intermediate tetrahedron must be of $C$-kind.
Thus, the degeneracy can be estimated as follows: $2^{N_{T}^\prime}<{\cal N}_{AC}<3^{N_{T}^\prime}$, 
where $N_{T}^\prime=N/4$ is the number of tetrahedra in one sublattice.
We conclude that the degeneracy of this manifold is still macroscopic, but smaller than that of $B$ one.

Thus, even though the formation of spin-singlets removes the spin degeneracy of the pyrochlore lattice,
there is still a  macroscopic degeneracy to be lifted.
The main question is whether this degeneracy can be removed by extending our effective Hamiltonian to the previously neglected $dd\pi$ and $dd\delta$ overlaps.
When these processes are considered, $b_2$ and $b_3$ bonds acquire a different bonding energy,
and, as number of these bonds is not the same for $B_i$ and $AC$ manifolds,\cite{prl} 
the degeneracy between these two phases  can be lifted.
However, this effect is smaller than the one induced by magnetoelastic coupling (see Sec. III), as 
$J_{dd\pi}\equiv \frac{t_{dd\pi}^2}{U_2}\simeq 2$ meV, while magnetoelastic energy gain per ion is about $6.5$ meV. It is then obvious to look for the degeneracy 
removal first in terms of this "spin-Teller" interaction, as done below.

Moreover, the degeneracy within the $B_i$-manifold can not be removed within the 
effective electronic model, not even introducing
smaller NN hopping integrals. 
The reason is related to the fact that the energy gain depends only on the total 
number of bonds of each type ($n_{b_0}$, $n_{b_1}$, $n_{b_2}$, $n_{b_3}$) in the unit cell, 
and, in order to fill the whole crystal 
with a periodicity not lower than the one of the primitive cubic cell,
the average number of bonds $n_{b_i}$ per tetrahedron should be  the same,
whichever of the three tetrahedral configurations is taken ($B_1$, with $n_{b_2}$=1 and $n_{b_3}$=2; or $B_2$, with $n_{b_2}$=3 and $n_{b_3}$=0; 
or $B_3$, with $n_{b_2}$=2 and $n_{b_3}$=1). 
This number is given by $n_{b_0}$=1, $n_{b_1}$=2,
$n_{b_2}$=2, $n_{b_3}$=1, and it corresponds to the value of $B_3$ case, that is the only 
one that allows to cover the whole cubic cell without mixing 
to other configurations (see Fig. \ref{vbc}).

From the above discussion it follows that only correlations between 
bonds can lift this degeneracy. 
These correlations naturally appear if the magneto-elastic contribution to
the energy is considered. The orbitally-driven modulations of the 
spin exchange interactions will distort the underlying lattice through the 
spin-Peierls mechanism and different distorted patterns will pay a different elastic energy.
The three degenerate phases which will be discussed in the next section
are those with the unit cell filled by all $B_3$ tetrahedra ("$B_3$-phase"), 
that with a mixture of $B_1$ and $B_2$ tetrahedra ("$B_1B_2$-phase"), 
and that where the unit cell is filled by alternated $A$ and $C$ tetrahedra 
("$AC$-phase").

\section{Evaluation of the magnetoelastic energy}

\begin{figure}
\epsfysize=22mm
\centerline{\epsffile{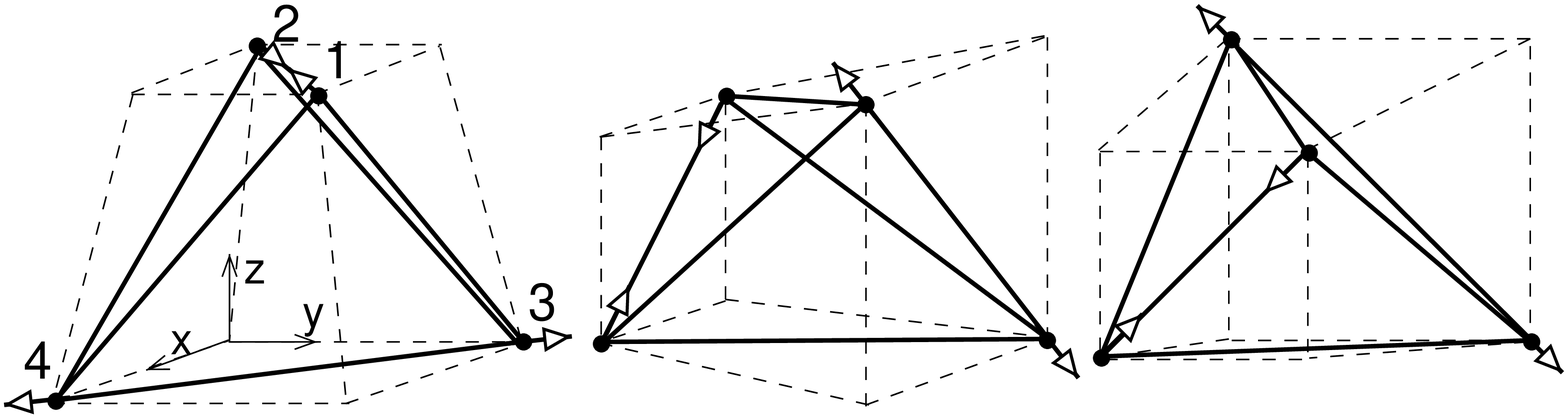}}
\vspace{0.5cm}
\caption{ 
 Triplet-T deformation mode from the irreducible  representations 
of the tetrahedron group. This mode generates a distortion
 of the tetrahedron, 
with short  and long bond  located opposite to 
each other and four intermediate (undistorted) bonds.}
\label{deform}
\end{figure}

In a spin-Peierls system the magnetic energy gain due to the spin-singlet pairs outweights the increase 
in elastic energy due to the dimerization of the regular array. We have shown 
with qualitative arguments\cite{prl} that in our case this mechanism can select the 
 triplet-T normal mode of the tetrahedron group, leading to a distortion with one short and one long bonds 
located at opposite edges and four undistorted bonds (see Fig. \ref{deform}).
Indeed, a reduction of the bond length increases the magnetic energy gain and therefore favors the shortening of the bond with the strongest superexchange, 
i.e., $b_0$, where the singlet is located.

Even though this picture is correct in its basic features, yet a quantitative description of the global physical mechanism leading to the tetragonal 
distortion requires a more careful analysis, that takes into account all elastic normal modes of the single tetrahedron, as well as the correlations 
of these normal modes within the unit cell.
The aim of the present section is just to analyze such a global mechanism.

The dependence of the energy gain $\Delta E$ on the magnitude of distortion can be evaluated  as a sum of magnetic $\Delta E^m$ 
and elastic $\Delta  E^{el}$ contributions at each bond: 
\begin{equation}
\Delta E =\sum _{ij}\Delta E_{ij}^{m}+\Delta E_{ij}^{el}.
\label{energy}
\end{equation}

The sum is restricted to the NN  sites.

All bonds in the undistorted lattice have the same length and we represent their elastic energy assuming that 
all ions are connected one another through equal springs of constant $k$: 
\begin{equation}
\Delta E_{ij}^{el}= \frac{1}{2} k (\delta d_{ij})^2=\frac{1}{2} C_0 \frac{(\delta d_{ij})^2}{d_0^2} 
\label{el}
\end{equation}
\noindent 
where the constant $C_0\equiv k d_0^2$ is the radial force constant\cite{harris} and
$\delta d_{ij}\equiv d_{ij}-d_0$ is the deviation  of the bond length from its value $d_0$ in the undistorted cubic lattice.

The magnetoelastic energy of the generic bond $ij$ can be written as:
\begin{equation}
\Delta 
E_{ij}^{m}=(J(d_0)+g\delta d_{ij})\vec {S}_i\cdot \vec {S}_j 
\label{magn}
\end{equation}
\noindent where $g\equiv \frac{\partial J(d)}{\partial d}|_{d=d_0}$. 
The dependence of exchange constants on the distance is mediated by the hopping matrix element: exchange constants are proportional to $t^2$, 
and $t$ is inversely proportional to the fifth power of the distance. In Ref. [\onlinecite{harris}] it is possible to find a rough estimate 
of the proportionality constant, $\alpha$: $t=\alpha /d^5$, with $\alpha =\frac{3}{4}\eta_{dd\sigma}\frac{\hbar ^2 r_d^3}{m}$. 
Here $r_d$ is a characteristic length, that for Ti
is 1.08 \AA, while $\eta_{dd\sigma}=-16.2$ and $\frac{\hbar ^2}{m}=7.62 eV\cdot$ \AA$^2$, giving the value $\alpha =-116.628$ eV$\cdot$\AA$^5$.
It is immediately apparent that even a small reduction in the length of a given bond can 
lead to a relatively high magnetic-energy gain.

In order to analyze the magnetoelastic behavior of the three degenerate phases of our spin-orbital model we first focus on a single tetrahedron, 
either of $A$, $B_i$ or $C$ type. For a more fluent reading, the detailed calculations are reported in Appendix B, and here we just comment on the main points.

\begin{figure}
\epsfysize=85mm
\centerline{\epsffile{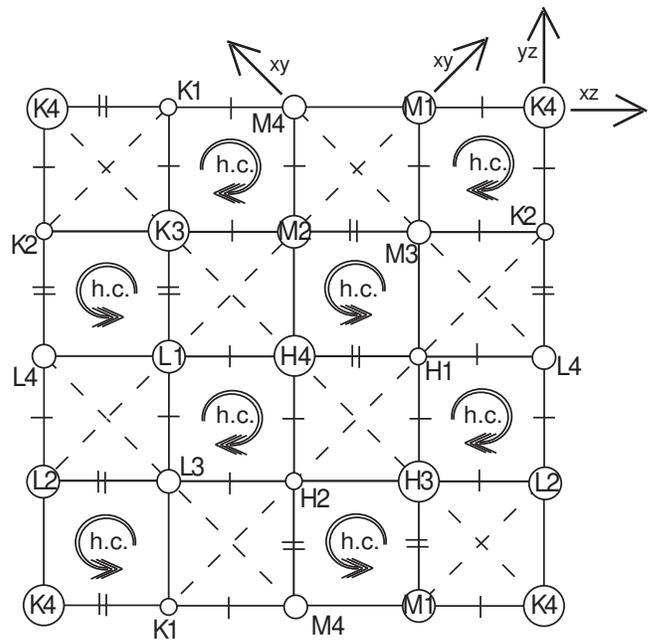}}
\caption{Bidimensional projection of the unit cell. Circles represent Ti-ions, with the same labels as in Fig. \ref{mixed}. 
Diameters are proportional to the $z$ coordinate of the ion:  $z=1,0.75,0.5,0.25$, in fractional units. 
All bonds are oriented along one of $xy$, $xz$, $yz$ bisectors, as indicated. Helical chains (h.c.) 
are represented in the direction of increasing z. Doubly cut bonds are $b_0$ bonds and singly cut are $b_1$ bonds.}
\label{plconstr}
\end{figure}

In general, the global magnetoelastic Hamiltonian involves more than one normal mode for each tetrahedron: for example, for $A$ tetrahedra 
the singlet and one doublet modes contribute to the magnetoelastic energy, and for $B_2$ tetrahedra all six normal modes are involved, 
as reported in Appendix B. Moreover, even though the triplet T-mode associated to the singlet bond leads to a bigger energy gain for $B_3$ 
than for $B_1$ tetrahedra, when all available normal modes are considered, the total energy gain gets different contributions from all normal modes, 
but in such a way that it is the same for all three $B_i$ tetrahedra, and equal to $-4.5\frac{g^2}{k}$. What makes the difference among 
these configurations in a specific case like that of MgTi$_2$O$_4$ is that when these tetrahedra are considered in an infinite lattice, 
the single-tetrahedron degeneracy is lifted by the correlations among tetrahedra. In fact, in correspondence with their common energy minimum, $B_3$, $B_1$, 
and $B_2$ tetrahedra are all distorted differently, as can be deduced from the location of their minima. Thus, when these distortions 
are combined together in order to fill in exactly the global cubic cell, as in Fig. \ref{plconstr}, the appearance of new constraints, due 
to the relative match of all tetrahedra within the cell, forces the solution to another minimum. For example, in order to keep the global 
volume unaltered, all bond distortions due to spin-singlets must appear with reversed sign from one tetrahedron to the other, because to each expansion  
there must be a corresponding contraction.
Notice that also the $AC$-phase is still degenerate with $B_i$ configurations, as far as just magnetoelastic energy gain associated with 
single tetrahedra is considered, as, in average, it is again $(-8-1)/2\frac{g^2}{k}=-4.5\frac{g^2}{k}$ (see Appendix B).
Remarkably, when inter-tetrahedra correlations are considered, elastic singlet modes contribute to $AC$-phase differently of $B_i$ phase. In fact, it is worthwhile for the system 
to expand $C$ tetrahedra and contract $A$ tetrahedra of an equal amount, as this leads to a net energy gain, because the singlet mode energy gain of $A$ tetrahedra exceeds that of $C$ ones.

It is clear from these considerations that in order to get the global magnetoelastic minimum for the unit cubic cell we cannot rely on the unconstrained 
minimization of the single tetrahedra shown in Appendix B. One alternative possibility is to calculate the magnetoelastic energy for the normal modes 
of the global cubic cell for the three degenerate configurations ($B_3$, $B_1B_2$, and $AC$-phase), in such a way as to automatically consider 
all constraints among tetrahedra. 
Yet, this procedure is not complete, as normal modes corresponding, e.g., to the buckling of tetrahedra do not contribute to the magnetoelastic energy. Such a contribution appears only when the normal mode involves an increase/decrease of some bond length, and, thus, 
it can be expressed as a linear combination of the normal modes of each single tetrahedron, with the constraint that the position of each 
equivalent Ti-ion over all nearest neighbor unit cells be the same (${\bf k}=0$ modes).


In order to fulfill this constraint, we impose that the 
total length over all the straight bonds connecting one ion with the equivalent one in the next cubic cells is a conserved quantity. 
This means that, for example, $d_{H4-H3}+d_{H3-K4}+d_{K4-K3}+d_{K3-H4}$ (see Fig. \ref{plconstr}) is a conserved quantity, and, thus, the sum of the four displacements $\delta d_{H4-H3}+\delta d_{H3-K4}+\delta d_{K4-K3}+\delta d_{K3-H4}=0$. The same happens for all other 3 chains in the $xy$-plane, for the 4 of the $xz$-plane, and for the 4 in the $yz$-plane. 
Due to Eqs. (\ref{normal2}), these twelve conditions become constraints for the normal variables of the different tetrahedra.
The search for the minima of the magnetoelastic Hamiltonians (\ref{meA}-\ref{meC1}) with the previous constraints leads to the following expression of the energy per unit cell: $-18\frac{g^2}{k}$ for $B_3$-phase; $-17.2\frac{g^2}{k}$ for $B_1B_2$-phase; $-14.4\frac{g^2}{k}$ for $AC$-phase. Thus, the ground state is given by $B_3$, with an energy per tetrahedron of $-2.25\frac{g^2}{k}$, corresponding to the sum of the energies of the triplet $t_2$ and doublet $e_1$ modes. The singlet and the doublet $e_2$ modes do not contribute to the energy, when they are constrained on the global unit cell. As now each ion is shared by two tetrahedra, the energy per site is $-1.125\frac{g^2}{k}$.

However, there is an experimental fact that is not properly explained by this solution, and for which an extra discussion is required. 
The solution of the constrained minimization gives an amplitude of $\frac{2g}{k}$ for triplet $t_2$ mode and $\frac{g}{k}$ for doublet $e_1$ mode, thus a 
ratio 2:1 in favor of the triplet mode.
But the experimentally reported tetrahedral distortions\cite{schmidt} are, referring to Eqs. (\ref{normal}), and (\ref{normal2}): $\delta r_{13}=-0.155$ \AA, 
$\delta r_{14}=-0.006$ \AA, which imply, when only $t_2$ and $e_1$ normal modes are active, $e_1=-0.012$ \AA, and $t_2=-0.149$ \AA, 
with a triplet-to-doublet ratio of more than 10.
This indicates that, in the previously outlined model, some features of the real system are not taken into account, like the elastic potential coming from the oxygens, 
that acts in such a way as to reduce some distortions.
Thus, the possibility to get better quantitative results for the experimental bond-lengths within our scheme should passes through a way to mimic this aspect 
by means of a stronger constraint on the available modes before the minimization procedure. 
If we consider Fig. \ref{plconstr}, we can see that the ion $H4$ can be connected to the equivalent ones in the neighboring edge-sharing unit cells through one of 
the three "linear" paths: $H4-H1-L4-L1-H4$, in $xz$-plane, or $H4-H2-M4-M2-H4$, in $yz$-plane, or $H4-H3-K4-K3-H4$, in $xy$-plane. 
Face-sharing unit cells are connected by the paths along the helical chains, 
namely: $H4-H2-L3-L1-H4$ and $H4-H1-M3-M2-H4$, for chains along $z$-axis, $H4-H2-K1-K3-H4$ and $H4-M2-M1-H3-H4$, for chains along $x$-axis, 
$H4-H1-K2-K3-H4$ and $H4-H3-L2-L1-H4$, for chains along $y$-axis. In principle the global $H4-H4$ distance along, e.g., the $z$ axis can be kept 
fixed even if the sum of the four bonds changes, as the bonds are not along a straight line, and the elongation of one bond can be combined with an 
appropriate rotation, due to a tetrahedron buckling. Yet, this buckling implies an extra elastic energy loss due to the relative movement with Mg and O ions. We can forbid these extra losses, by imposing another series of constraints that express the global conservation of the total length also along the "helical" chains, in this way freezing some of the normal modes that were previously allowed.

When the minimization is performed with this new constraint, the results for the energy per unit cell are: $-16\frac{g^2}{k}$ for $B_3$-phase; 
$-13\frac{g^2}{k}$ for $B_1B_2$-phase; $-\frac{g^2}{3k}$ for the $AC$-phase. Again, $B_3$-phase is the lowest, but this time the amplitude of 
$e_1$ mode is zero and the minimum energy is reached for a pure triplet mode, of amplitude $\frac{2g}{k}$. Notice that with this constraint the $AC$-phase gains very few, and also the relative energy difference between $B_1B_2$-phase 
and $B_3$ increases.

Finally, let us stress again that the main result is that when correlations among magnetoelastic energy are taken into account, $B_3$ phase is always 
stabilized with respect to the two configurationally degenerate $B_1B_2$ and $AC$ ones. Of the two criteria used for preserving the cell volume, the second seems more in keeping with experimental data. This suggests that oxygens play a role in the elastic distortions of the spinel cell, which is more rigid towards bucklings than the bare pyrochlore lattice. 
In any case it is found that the elongation of the non-interacting $b_{3}$ bond and the corresponding stretching of $b_0$ bond throughout the whole cell is
 always energetically more favorable, as found experimentally.

Given the previous $d$-dependence of the hopping amplitude $t$, it is possible to estimate the gain energy per ion associated with this magnetoelastic distortion. 
First, $g\equiv -\frac{\partial J}{\partial d}|_{d=d_0}=10 \frac{J}{d_0}$. Using Eq. (\ref{normal2}), $\delta r_{13}=-0.155$ \AA$=-\frac{2g}{k}$,
 with $k=\frac{C_0}{d_0^2}$. This implies that $C_0=\frac{10Jd_0}{0.155}\simeq 9.7$ eV, if we take $d_0\simeq 3.008$ \AA. Finally, the estimate of 
the average magnetoelastic energy per site can be easily found: there are 16 ions in one unit cell, and thus the energy per site is 
$E_{me}= -\frac{g^2}{k}=\frac{(10J)^2}{C_0}\simeq 6.5$ meV.

\section{Effect of Magnetic field}

As we have shown above the ground state of the system is non-magnetic and spins are paired in 
singlet states. The spin degrees of freedom are thus gapped and the gap of triplet excitations 
is given by the singlet binding energy $\Delta_s=4J[1-2\eta]$ (see Eq. (\ref{b0})).
When an external magnetic filed is applied, the energy of triplet excitations decreases,
due to the gain of Zeeman energy, while the singlet state does not experience the applied field.
Therefore, one would expect no change in the symmetry of the ground state 
up to fields $g\mu_B H_c=\Delta_s$ at which singlet triplet gap closes 
($g$ is the gyrotropic factor and $\mu_B$ the Bohr magneton).
At $H=H_c$ a second order transition can in principle take place from non-magnetic
to a magnetically ordered state, driven by condensation of lowest energy triplets at
some ordering wave-vector ${\bf Q}$ at which triplet spectrum has a  minimum.

However, the presence of an orbital degeneracy can modify this convectional picture.
As shown in  Sec. II, the first excited state above Dimer phase is
the FM one, characterized by $C$ orbital pattern. 
The energy of FM phase, contrary to the Dimer phase, is sensitive to the applied magnetic field and decreases of $\Delta E_H=-gH\mu_{B}S_{i}^{z}$.
The critical field ${\tilde H}_{c}$ at which the energies of the two phases are equal is:
\begin{equation}
g\mu_B{\tilde H}_c=J[2-11\eta],
\label{hc}
\end{equation}
and one can easily verify that ${\tilde H}_{c}< H_c$.
Therefore at $H={\tilde H}_c$ there will be a first order transition from Dimer phase to FM state.
This transition will be accompanied by a simultaneous rearrangement of orbital ordering 
and the abrupt closure of the spin gap, as the FM phase is gapless.
We can estimate the order of magnitude of ${\tilde H}_c$, 
for $\eta\simeq 0.15$ and $J=25$ meV, as $\mu_B {\tilde H}_c \simeq 4.4$ meV, i.e., 
a critical field of ${\tilde H}_c\simeq 76$ Tesla. 
In this estimate we did not consider the extra stabilization energy 
in favor of the singlet phase coming from the magnetoelastic distortion. 
Of course, one can read Eq. (\ref{hc}) the other way round, 
in terms of $\eta$, and deduce that, if a critical field is found at a lower value than $76$ T, 
this implies that $J_H/U_2$ is closer to the critical value 
of $0.18$ than what estimated here.

\section{Detection of orbital ordering through RXS.}

As already outlined in the previous two sections, orbital ordering in MgTi$_2$O$_4$ is mainly dictated by superexchange interactions. 
This implies that the orbital orientation at each site does not strictly follow the symmetry imposed by the local crystal field, 
as happens when the effect is purely determined by Jahn-Teller distortions.
Therefore, we believe that there is the possibility to exploit the local symmetry differences in the helical d$_{xz}$-d$_{yz}$ 
orbital pattern by means of resonant x-ray scattering (RXS), where the local transition amplitudes are added with a phase factor
 that can compensate the vanishing effect due to the global tetragonal symmetry.
Conceptually, this procedure is the analogue of the one used in the case of manganites\cite{murakami} that led to a series 
of results initially interpreted as a direct evidence of orbital ordering,\cite{maekawa} and soon 
later recognized\cite{elfimov,natoli,igarashi} as mainly determined by the oxygen distortion around each Mn-ion. Yet, in this case Bragg-forbidden reflections (e.g., for LaMnO$_3$\cite{murakami}) relate two sites with different orbital occupancy {\it and}
 local distortion. Thus, at K edge, the signal at a given energy turns out to depend on the difference between, e.g., $p_x$ and $p_y$ density of states
 around that energy,\cite{note} projected on Mn-ions. Such local anisotropies in the electronic $4p$ density of states can be a consequence 
of oxygen distortions (Jahn-Teller effect) or can be induced by the ordering of the underlying $3d$ orbitals through the
 $3d-4p$ Coulomb repulsion $U_{dp}$ (orbital ordering effect).\cite{natoli}
The results of Refs. \onlinecite{elfimov,natoli,igarashi} independently show that the effect of the Jahn-Teller mechanism is
 much stronger, about a factor 10 in amplitude, than the effect of orbital ordering. This led to the conclusion that OO cannot
 be "directly" probed by means of RXS at K edges, where the strength of the ligand field overwhelms the effect induced by $U_{dp}$. 
Due to this result, the search for experimental evidence of orbital ordering moved to L edge RXS\cite{altarelli} where one is sensitive directly to $3d$ orbitals.
Yet, this kind of spectroscopy has the serious drawback that, in order to have non-imaginary Bragg angles, it is necessarily 
limited to crystals with very big unit cells, and this is not feasible for MgTi$_2$O$_4$. In fact, Ti L$_{2,3}$ edges are 
at about 453.8 and 460.2 eV, which corresponds to a photon wavelength $\lambda \simeq 27$ \AA, that does not allow Bragg 
law $\sin \theta_B = \lambda/2d$ to be verified, not even for $(001)$ reflection.

In spite of all this, we believe that it is still possible to detect the proposed OO at K edge. The key-point to go beyond 
the conclusions of Refs. [\onlinecite{elfimov,natoli,igarashi}] lies in the fact that their results are strongly related to 
symmetry and distortion of the system as well as the kind of reflection under study. In particular, while the signal induced 
by the Coulomb repulsion $U_{dp}$ is of the order of $\simeq 200\div 300$ meV, independently of the particular crystal structure,
\cite{nota4} the influence of the oxygen distortion is much lower in the directions of $t_{2g}$ orbitals than in $e_g$ ones. 
It might indeed seem conceivable that a reduced Jahn-Teller distortion can give rise to a sizable interference with the 
OO-induced effect in the RXS intensity. Moreover, there is the possibility that the rotation of the exchange vector and/or 
polarizations in the RXS experiment can make the effect more or less pronounced, according to the experimental conditions.
We have thus performed a numerical simulation and found that it is indeed possible to experimentally reveal the presence of 
the OO through the comparison of different RXS signals.
In the following we illustrate the details of our calculations, in order to clarify the previous theoretical speculations.


The transition process for RXS, governed by the Fermi Golden rule, depends on the state overlap through the polarized electric field of the incoming ($i$)
 or outgoing ($o$) photon.
If we consider the multipole expansion up to the electric quadrupole contribution, we get:

\begin{equation}
M_{ng}^{i(o)} = \langle \psi_n|\vec{\epsilon}^{i(o)} \cdot \vec{r} ~ \big(1 - \frac{1}{2}i\vec{k}^{i(o)} \cdot \vec{r}\big) |\psi_g \rangle
\label{eq_m}
\end{equation}

Here $\psi_g$ and $\psi_n$ are ground and intermediate state wave functions, respectively, $\vec{\epsilon}^{i(o)}$ is the polarization of the
incoming (outgoing) photon and  $\vec{k}^{i(o)}$ its wave vector. Around an absorption edge, $M_{ng}$ is highly energy and angular dependent 
and this provides the sensitivity to the electronic structure around the atom. 
In RXS the global process of photon absorption, virtual emission of the photoelectron, and subsequent decay with re-emission of a photon, 
is coherent, thus giving rise to the usual Bragg diffraction condition. The outgoing photon can have different
polarization and wave vector compared to the incoming one and the
atomic anomalous scattering factor (ASF) reads\cite{blume}:

\begin{equation}
f=f'+if'' = \frac{m_e}{\hbar^2}\frac{1}{\hbar\omega} \sum_{ng}
\frac{(E_n-E_g)^3M_{ng}^{o*}M_{ng}^i}
{\hbar\omega-(E_n-E_g)-i\frac{\Gamma_n}{2}} \label{eq_arxs}
\end{equation}

Here $\hbar\omega$ is the photon energy, $m_e$ the
electron mass, $E_g$ and $E_n$ are the ground and 
intermediate state energies and $\Gamma_n$ is the broadening of
the  transition. The sum over the intermediate states
starts from the Fermi energy.\cite{yvesprb}

In the tetragonal phase of MgTi$_2$O$_4$ the space group is the chiral group P$4_12_12$, No. 92 of Ref. [\onlinecite{tables}] 
(or its "mirror"-related P$4_32_12$, No. 96), with 8 ions per unit cell, and the structure factor is:

\begin{equation} 
A(\vec{Q})=\sum_{j=1}^8 e^{i\vec{Q}\cdot\vec{R}_j}f_j
\label{structfac}
\end{equation}

\noindent where $f_j$ is the ASF of ion $j$. The atomic positions, their orbital filling and the symmetries relating the ions one another are schematized in Table I. 
Notice that the tetragonal cell is characterized by a 45-degree rotation in xy-plane, compared to the cubic cell of Fig. \ref{fig2}(a).

\begin{eqnarray*}
\begin{array}{c|cccc}
 {\rm atom} &  {\rm OO} & {\rm position} & {\rm symmetry} \\
\hline
{\rm Ti}_1 (H3) &  d_{yz}  &  (u,v,w)   &  {\hat E}   \\
{\rm Ti}_2 (L4) &  d_{yz}  & (-u,-v,\frac{1}{2}+w) &  {\hat C}_{2z}    \\
{\rm Ti}_3 (H1) &  d_{xz}  & (\frac{1}{2}-v,\frac{1}{2}+u,\frac{1}{4}+w) & {\hat C}_{4z}^+   \\
{\rm Ti}_4 (L2) &  d_{xz} & (\frac{1}{2}+v,\frac{1}{2}-u,\frac{3}{4}+w)  & {\hat C}_{4z}^- \\  
{\rm Ti}_5 (L3) & d_{xz} & (\frac{1}{2}-u,\frac{1}{2}+v,\frac{1}{4}-w) & {\hat C}_{2y}  \\
{\rm Ti}_6 (H4) & d_{xz} & (\frac{1}{2}+u,\frac{1}{2}-v,\frac{3}{4}-w) & {\hat C}_{2x}  \\
{\rm Ti}_7 (H2) & d_{yz} & (v,u,-w) &  {\hat C}_{2xx}  \\
{\rm Ti}_8 (L1) & d_{yz} & (-v,-u,\frac{1}{2}-w) & {\hat C}_{2{\overline{x}}x} 
\end{array}
\end{eqnarray*}
\begin{center}
TABLE I. {\small Orbital occupancy and tetragonal position of Ti-ions. Symmetry operations are referred to Ti$_1$.}
\end{center}

Here $u = 0.9911, \; v = 0.2499, \; w = 0.8668$ are the fractional
coordinates of the atoms in unit of tetragonal axes. 
Notice that such a derivation allows also to deal with x-ray natural circular dichroism (XNCD), as the absorption cross section corresponds 
to the imaginary part of the forward scattering amplitude (i.e., when $\vec{Q}=0$). Indeed a signal in the dipole-quadrupole interference channel
 could be expected, as the space symmetry group of MgTi$_2$O$_4$ is a chiral one, and there could be the possibility that it is affected by OO. 
Unfortunately, our numerical simulations with the finite difference method,\cite{yvesprb} in the same range of parameters as shown below for RXS, 
demonstrates that such a dichroic signal is too low (less than 0.05 \% of the absorption) and that the changes induced by OO are practically negligeable.

For this reason in the following we shall analyze just K edge RXS, and in particular all reflections $(00l)$, that are Bragg-forbidden unless $l=4n$.
Using the symmetries of Table I, the structure factor can be expressed in the following form:

\begin{eqnarray}
A(00l)=(1+(-)^l {\hat {C}}_{2z}) (1+(i)^l {\hat {C}}_{4z}^+) \nonumber \\
(e^{2\pi i l w}+(i)^l e^{-2\pi i l w} {\hat {C}}_{2y})f_1
\label{eq_t}
\end{eqnarray}

Note that even if $f_1$ is a scalar, it is expressed as a scalar product of two tensors, and, in Eq. (\ref{eq_t}), rotation operators 
are understood as applied to one of them. We can limit to  the dipole approximation, where we just need to consider rank-2 cartesian tensors,
 $T_{\alpha\beta}$, whose irreducible representation in SO(3) are a scalar, a pseudovector and a traceless symmetric tensor: the first never
 contributes to Bragg-forbidden reflections (which are scalar-forbidden), while the second is proportional to a magnetic moment and has no 
influence on non-magnetic systems like MgTi$_2$O$_4$. The third tensor has 5 independent components, that can be labeled according to the
 usual second-rank real spherical harmonics as: $D_{xy}$, $D_{xz}$, $D_{yz}$, $D_{x^2-y^2}$, and $D_{3z^2-r^2}$. They are a measure of the 
anisotropy of $p$-density of states projected on the resonant ion.\cite{dimasync} For example, $D_{x^2-y^2}$ measures the difference in the 
density of states in x-direction, $p_x$, and in y-direction, $p_y$.
In order to determine the implications of Eq. (\ref{eq_t}) on the various reflections, we need to know how the symmetry operators act on our 
symmetric tensors. The indices of each tensor change according to the following rules: 
${\hat {C}}_{2z} (x,y,z) = (-x,-y,z)$; ${\hat {C}}_{2y}(x,y,z) = (-x,y,-z)$ and ${\hat {C}}_{4z}^+(x,y,z) = (y,-x,z)$. 
Similar results are obtained reminding that the ASF in our approximation is a second-rank irreducible tensor $f^{(2)}_m$, 
and ${\hat {C}}_{2z}f^{(2)}_m=(-)^m f^{(2)}_m$, ${\hat {C}}_{4z}^+ f^{(2)}_m=i^m f^{(2)}_m$, and ${\hat {C}}_{2y}f^{(2)}_m=(-)^m f^{(2)}_{-m}$.
As a consequence of Eq. (\ref{eq_t}) we get that all Bragg-forbidden reflections of the kind (0,0,4n+1) and (0,0,4n+3) are sensitive 
to the complex mixture of density of states: $D_{xz}\pm iD_{yz} \propto Y^2_{\pm 1}$. For this reason, due to the cylindrical symmetry, 
a constant azimuthal scan is expected as well as a minor dependence on the orbital anisotropies of $xz$ and $yz$ kind.  
On the contrary, Bragg-forbidden reflections of the kind (0,0,4n+2), if we neglect $\cos 4\pi w \simeq 0.1\ll \sin 4\pi w \simeq 1$, 
are just proportional to $D_{xy}$: thus, they must show a non-constant azimuthal scan, and, indirectly, manifest some properties related 
to the OO, through the depletion of the filled $d_{xy}$ orbitals.

\begin{figure}
\centerline{\epsfig{file=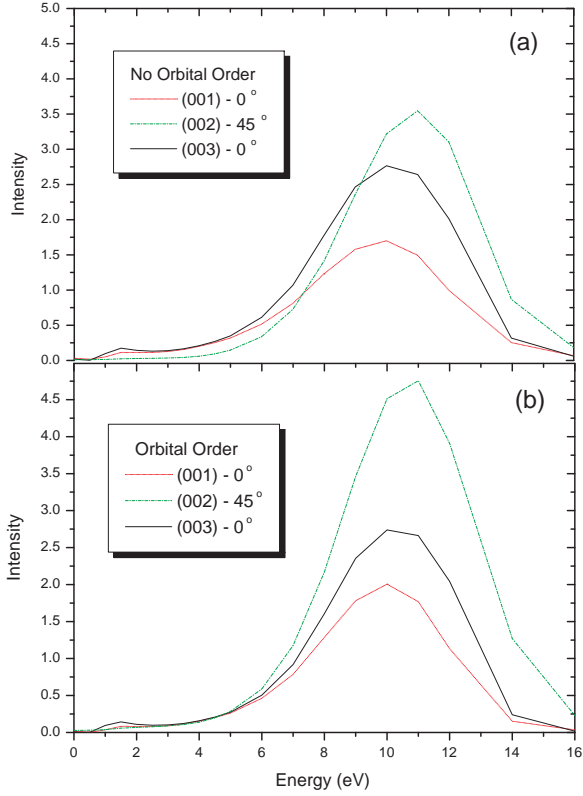,width=10cm}}
\vspace{-1cm}
\caption {RXS intensities for $(00l)$ reflections ($l=1,2,3$) without OO (a) and with OO (b). Angles refer to incoming $\sigma$ polarization: 
$\phi=0^o$ coincides with the cubic x-axis and $\phi=45^o$ with the tetragonal x-axis. The cluster radius is $3.3$ \AA.  The reference energy is the Fermi level.}
\label{figrxs}
\end{figure} 

If we express the polarization dependence in the various channels, referring to the cubic frame of Fig. 1, we get the following results. 
Both $(001)$ and $(003)$ reflections couple to $\epsilon_x\epsilon_z$ and $\epsilon_y\epsilon_z$, and thus they are different from zero only in the $\sigma\pi$-channel.
As their amplitude is proportional to $(\sin\phi+i\cos\phi)$, they have a constant azimuthal scan.
Instead, $(002)$ reflection is detectable in all $\sigma\sigma$, $\sigma\pi$ and $\pi\pi$ channels, with an intensity that scales, respectively, 
as $\sin^2(2\phi)D_{xy}^2$, $\sin^2\theta_B\cos^2(2\phi)D_{xy}^2$, $\sin^4\theta_B\sin^2(2\phi)D_{xy}^2$, where $\theta_B$ is the Bragg angle.
 As $\sin^2\theta_B\simeq 0.1$, signals in $\pi\pi$-channel are 1/10 smaller than those in $\sigma\pi$ and 1/100 smaller than those in $\sigma\sigma$-channel.
 Also their azimuthal scans are out of phase: that of $\sigma\pi$ channel has its maximum value when $\phi=0$, i.e., in the direction of the nearest neighbor oxygens,
 while those of $\sigma\sigma$ and $\pi\pi$ channels take their maximum value along the Ti-Ti chains in the xy-plane, where $\sigma\pi$ signal is zero. 

In Fig. \ref{figrxs} we show the results of our numerical simulations for $(00l)$ reflections, performed with the finite difference method option of 
the FDMNES program.\cite{yvesprb} As a input file we used the refined positions of Ref. [\onlinecite{schmidt}] for Mg, Ti and O ions. Because of the
 increasing CPU-time consuming, we performed our simulation, with and without orbital ordering, just for a cluster of radius 2.5 \AA, and one of radius
 3.3 \AA, the first including the oxygen octahedron and the second also all 6 nearest neighbors Ti-ions around the resonant atom. All assertions stated before,
 regarding the relative intensities and polarization dependence, have been confirmed numerically.

The main result of our numerical simulations lies in the comparison of Fig. \ref{figrxs}(a) and Fig. \ref{figrxs}(b), that shows how to reveal the presence 
of OO through a simple experiment. In fact, the OO case is characterized by a big increase in intensity of $(002)$ reflection, of a factor of 1.7, 
while $(003)$ is left unaltered and $(001)$ increases of no more than 1.3. Thus, an experimental setup that measures these signals keeping track of 
the relative intensity is sufficient to determine whether the orbital order is present or not: if the three reflections have almost the same intensity, 
there is no OO, while if the $(002)$ is about twice than the others, this is the signature of the OO of the kind predicted by our theory.

\begin{figure}
 \centerline{\epsfig{file=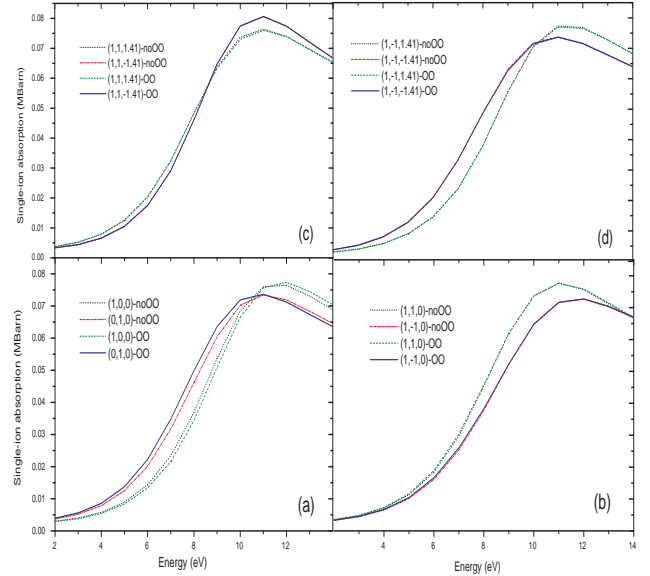,width=10cm,height=10cm}}
\caption {Absorption spectra for a TiO$_6$ cluster in MgTi$_2$O$_4$. Directions of incident polarizations are expressed in the tetragonal frame.
 Energy zero is at the Fermi level. For each direction the comparison is made between the OO case and that without OO.}
\label{xansing}
\end{figure} 

In order to understand the reasons behind this enhancement in intensity, it is useful to study K edge absorption, with and without OO, in the fictitious 
system where the only absorption center is Ti$_1$-ion. In fact, in this way we probe directly the $p$ density of states projected on Ti$_1$-ion, i.e., $f_1$, 
the square of which is the RXS intensity shown in Fig. \ref{figrxs}.
Thus, linear dichroism, e.g., along (100) and (010) tetragonal directions is sensitive to $D_{xy}$. In Fig. \ref{xansing} we report the absorption for a 
single Ti-ion surrounded by the oxygen octahedron.
Oxygens in $xy$-plane almost lie along cubic $x$ and $y$ axes, slightly squeezed towards tetragonal $x'$ axis. Labels in Fig. \ref{xansing} refer to the 
tetragonal system, so that (100) direction corresponds to the incoming polarization along the line connecting the nearest neighbor Ti-ions in xy-plane 
(45$^o$ from the x-axis of Fig. \ref{fig2}). It is evident that in all figures, except Fig. \ref{xansing}(a), the presence of OO does not change the
 signal and the dichroism is entirely due to the ligand distortion. In particular, Fig. \ref{xansing}(c) shows the dichroism in the $yz$ cubic plane, 
while Fig. \ref{xansing}(d) is taken in $xz$ plane and Fig. \ref{xansing}(b) is a measure of $p_x-p_y$, towards the oxygens.
On the contrary, Fig. \ref{xansing}(a) shows, in correspondence to its flex, an offset between the two curves varying from about 0.6 eV, 
in the absence of OO, to 0.9 eV, when OO is present. 

Finally, we want to comment about the influence of the cluster radius on the results. A series of numerical simulations of the local absorption 
(like those of Fig. \ref{xansing}) has been performed with radii $2.5$, $3.3$ and $4.0$ \AA, by means of the muffin tin program. 
The results show that the maximum dichroism between the (100) and (010) absorptions due to Jahn-Teller corresponds to the $3.3$ radius, 
where, compared to the $2.5$ cluster, two Ti-ions are added along the (100) tetragonal direction, whereas no Ti ions are present along the (010).
 The simulations with the $4.0$ \AA$~$ cluster shows again a reduction of dichroism compared to $3.3$ \AA, due to the cell composition. 
Thus, the results shown in Fig. \ref{figrxs} for the cluster radius of $3.3$ \AA$~$ correspond to the theoretical lowest ratio between the 
OO signal and JT one, and, thus, the effect of OO is expected to be not lower than that shown in Fig. \ref{figrxs}(b).


\section {Acknowledgments.}

Several stimulating discussions with C.R. Natoli and C. Lacroix are gratefully acknowledged.
N.B.P. acknowledges the hospitality of MPIPKS, Dresden, where part of the work has been performed.

\appendix

\section{The "full" effective Hamiltonian.}

In a previous Letter\cite{prl} we have treated the effective 
Hamiltonian \`a la Kugel-Khomskii\cite{kugel} with $dd\sigma$ contributions only.
In this appendix, for completeness, we take into account also the smaller terms proportional to the orbital overlaps of $dd\pi$ and $dd\delta$ type.
In this case the hopping matrix is no more diagonal as in Eq. \ref{orbpart}.
If we label planes and orbitals directions  as ${yz}\equiv 1$,  ${xz}\equiv 2$ and  
${xy}\equiv 3$, the matrix elements can be expressed in the form, given by Slater and 
Koster\cite{slater}:

\begin{eqnarray}
&&t_{ij(a)}^{aa}=3/4 dd\sigma+1/4 dd\delta \nonumber \\
&&t_{ij(a)}^{bb}=1/2 dd\pi + 1/2 dd\delta \label{hopping} \\
&&t_{ij(a)}^{ab}=0 \nonumber \\
&&t_{ij(a)}^{bc}=1/2 dd\pi - 1/2 dd\delta \nonumber
\label{hoppingb}
\end{eqnarray}

\noindent where $a$, $b$ and $c$ can be 1, 2, or 3. 
Lower indices denote the plane $(a)$ where the bond $ij$ lays, and upper ones label the 
orbitals. For example,  $t^{12}_{ij(3)}$ is the hopping matrix element from $yz$ to $xz$ orbitals along the bond $ij$ in $xy$-plane. 

Then, we introduce on-site transition orbital operators $T^{ab}_i$,
which describe initial and final orbital states in the hopping process at site $i$. 
If there is no change in the orbital state at site $i$,
then operators $T^{aa}_i$ are simply orbital projectors and can be written as:
$T^{11}_i=-\frac{1}{2}\tau_i^z (1-\tau_i^z)$, $T^{22}_i=\frac{1}{2}\tau_i^z (1+\tau_i^z)$ and 
$T^{33}_i = (1+\tau_i^z)(1-\tau_i^z)$. 
Non-diagonal matrix elements of the transition orbital operator describe
changes in the orbital state during the superexchange process: 
$T_i^{12}=\frac{1}{2}\tau_i^+\tau_i^+$,  $T_i^{21}=\frac{1}{2}\tau_i^-\tau_i^-$, 
$T_i^{13}=-\frac{1}{\sqrt{2}}\tau_i^+\tau_i^z$,  
$T_i^{31}=-\frac{1}{\sqrt{2}}\tau_i^z\tau_i^-$,
$T_i^{23}=\frac{1}{\sqrt{2}}\tau_i^-\tau_i^z$,  $T_i^{32}=\frac{1}{\sqrt{2}}\tau_i^z\tau_i^+$.

With these premises, the effective 
spin-orbital Hamiltonian can be written as:

\begin{eqnarray}
&&H_{\rm eff}=  
-\frac{1}{U_2-J_H} 
\sum_{\langle ij\rangle} {\big [}\vec S_i\cdot \vec S_{j} +3/4 {\big
  ]}O_{ij}^S
\nonumber\\
&&
+\frac{1}{U_2+J_H} \sum_{\langle ij\rangle} {\big [}\vec S_i\cdot \vec S_{j} 
-1/4 {\big ]}O_{ij}^S
\label{spinorb2}
\\
&&
+\sum_{\langle ij\rangle} {\big [}\vec S_i\cdot \vec S_{j} 
-1/4 {\big ]}(D_1O^{D_1}_{ij}+D_2O^{D_2}_{ij})
\nonumber
\end{eqnarray}

\noindent where the sum is restricted to the NN sites of the pyrochlore lattice.
The first term in $H_{\rm eff}$ describes ferromagnetic  interactions, while  second and the third terms  are antiferromagnetic. 
The explicit form of orbital contributions depends on the  kind of 
intermediate excited states involved in the superexchange process:
when the intermediate states have just singly-occupied orbitals ($O^S_{ij}$), their expression is:

\begin{eqnarray}
O_{ij}^S = 
\sum^3_{a=1}\sum^3_{b=1}\Bigl[
\sum^3_{\stackrel{c,d=1}{c\neq d}} 
t_{ij}^{ac}t_{ji}^{cb} T_{i}^{ab}T_{j}^{dd}+\sum^3_{\stackrel{c=1}{c\neq b}} 
 t_{ij}^{ac}t_{ji}^{ba} T_{i}^{aa}T_{j}^{bc}\Bigr] \nonumber \\
+\sum^3_{\stackrel{a,b=1}{a\neq b}} \sum^3_{\stackrel{c,d=1}{c\neq d}} 
t_{ij}^{ad}t_{ji}^{cb} T_{i}^{ab} T_{j}^{cd} ~~~~~~~~~~~~~~~~~~~~~~
\label{ondF}
\end{eqnarray}

This orbital term is common to both FM and AFM bonds, whose difference resides in their  energy 
denominators, $1/(U_2-J_H)$ for the FM bond and $1/(U_2+J_H)$ for the AFM.

The third term of the Hamiltonian describes the antiferromagnetic superexchange evolving through an 
intermediate excited state with double occupancy. The particular form with two energy denominators ($D_1$ and $D_2$) and 
two corresponding orbital terms is originated by the fact that orbitally doubly-occupied states are not eigenstates of the
 on-site multiband Hubbard Hamiltonian,\cite{v2o3} due to interorbital double hopping.
Exact eigenstates with single-orbital double occupancy are just linear combinations (with plus or minus sign) of two orbitally doubly-occupied 
ionic states, and have two different energies, $U_2+J_H$ and $U_2+4J_H$.
If, in the superexchange process, the electron forms a double occupancy 
at some orbital on site $j$ and comes back from the same orbital,   
then the energy denominator is $D_{1}=\frac{2}{3(U_2+J)}+\frac{1}{3(U_2+4J)}$. If, on the contrary,
the orbital state on site $j$ changes during the hopping process, then the energy denominator is
$D_{2}=-\frac{1}{3(U_2+J)}+\frac{1}{3(U_2+4J)}$.
The corresponding orbital contributions to $H_{\rm eff}$ are:

\begin{eqnarray}
O^{D_1}_{ij}=
\sum^3_{a,b,c=1}t_{ij}^{ab}t_{ji}^{bc} T_i^{ac} T_j^{bb} 
\label{Ddiag}
\end{eqnarray}

\begin{eqnarray}
O_{ij}^{D_2} =
\sum^3_{a,b=1} 
\sum^3_{\stackrel{c,d=1}{c\neq b}} 
t_{ij}^{ab}t_{ji}^{cd} T_i^{ad}T_j^{bc}
\label{Dnondiag}
\end{eqnarray}

In the absence of $dd\pi$ contributions, the hopping matrix connects only orbitals of the same type, $a$, which have nonzero overlap 
only in the corresponding $a$ plane. 
Then the  effective spin-orbital Hamiltonian (\ref{spinorb2}) simplifies to the  one we have obtained in Ref. \onlinecite{prl} 
and which we deal with in Section II.

\section{Magnetoelastic normal modes for a single tetrahedron.}

The normal modes of a single tetrahedron are a well-known argument: yet, as we need their 
explicit expressions both here and in Sec. III, we prefer to review the basic results.
Considering the small ionic displacements around the equilibrium positions, the coordinate of the 4 ions, when the origin is at the center of the tetrahedron, read:
$i_1\equiv (d_0+x_1,d_0+y_1,d_0+z_1)$, $i_2\equiv (-d_0+x_2,-d_0+y_2,d_0+z_2)$, $i_3\equiv (-d_0+x_3,d_0+y_3,-d_0+z_3)$, 
$i_4\equiv (d_0+x_4,-d_0+y_4,-d_0+z_4)$. Assuming that the four ions are connected each other with six equal springs of constant $k$, 
there will be an associated potential energy expressed as a bilinear form of the 12 displacements $x_1,...,z_4$. 
The diagonalization of this form leads to the 12 normal modes of the tetrahedron, six of which are the zero-energy modes associated 
to the three global translations and the three global rotation, that we may neglect in the following.


The remaining six eigenvectors can be classified into a triplet T-mode, with eigenvalue $k_t=4k$, a doublet E-mode, with eigenvalue $k_e=2k$ and a singlet S-mode,
 with eigenvalue $k_s=8k$. They all correspond to a particular deformation of the tetrahedron: those of T-modes are shown in Fig. \ref{deform}; 
those of the E-mode correspond to a local tetragonal or orthorhombic distortion (see, e.g., Ref. \onlinecite{tcher}), and that of the S-mode is the "breathing" mode, 
i.e., a volume expansion or contraction that does not change the tetrahedron shape.
If we introduce the stretches of each bond, $\delta d_{ij}$, in terms of the ionic displacement coordinates, 
it is then possible to express normal coordinates in terms of these latters.

\begin{eqnarray}
\delta d_{12}=\frac{1}{\sqrt{2}}(x_1-x_2+y_1-y_2) \nonumber \\
\delta d_{34}=\frac{1}{\sqrt{2}}(x_4-x_3+y_3-y_4) \nonumber \\
\delta d_{13}=\frac{1}{\sqrt{2}}(x_1-x_3+z_1-z_3) \nonumber \\
\delta d_{24}=\frac{1}{\sqrt{2}}(x_4-x_2+z_2-z_4)  \\
\delta d_{14}=\frac{1}{\sqrt{2}}(y_1-y_4+z_1-z_4) \nonumber \\
\delta d_{23}=\frac{1}{\sqrt{2}}(y_3-y_2+z_2-z_3) \nonumber 
\label{stretch}
\end{eqnarray}

\noindent and

\begin{eqnarray}
\theta_s&=&\frac{1}{2\sqrt{6}}(\delta d_{12}+\delta r_{34}+\delta d_{13}+\delta d_{24}+\delta d_{14}+\delta d_{23})       \nonumber \\
\theta_{e1}&=&\frac{1}{2}(\delta d_{13}+\delta d_{24}-\delta d_{14}-\delta d_{23})\\
\theta_{e2}&=&\frac{1}{2\sqrt{3}}(2\delta d_{12}+2\delta d_{34}-\delta d_{13}-\delta d_{24}-\delta d_{14}-\delta d_{23})\nonumber \\
\theta_{t1}&=&\frac{1}{2}(\delta d_{12}-\delta d_{34})  \nonumber \\
\theta_{t2}&=&\frac{1}{2}(\delta d_{13}-\delta d_{24})  \nonumber \\
\theta_{t3}&=&\frac{1}{2}(\delta d_{14}-\delta d_{23})    \nonumber 
\label{normal}
\end{eqnarray}

The inverted relations are:

\begin{eqnarray}
\delta d_{12}&=&\frac{2}{\sqrt{6}}\theta_s+\frac{1}{\sqrt{3}}\theta_{e2}+\theta_{t1} \nonumber \\
\delta d_{34}&=&\frac{2}{\sqrt{6}}\theta_s+\frac{1}{\sqrt{3}}\theta_{e2}-\theta_{t1} \nonumber \\
\delta d_{13}&=&\frac{2}{\sqrt{6}}\theta_s-\frac{1}{2\sqrt{3}}\theta_{e2}+\frac{1}{2}\theta_{e1}+\theta_{t2} \nonumber \\
\delta d_{24}&=&\frac{2}{\sqrt{6}}\theta_s-\frac{1}{2\sqrt{3}}\theta_{e2}+\frac{1}{2}\theta_{e1}-\theta_{t2} \\
\delta d_{14}&=&\frac{2}{\sqrt{6}}\theta_s-\frac{1}{2\sqrt{3}}\theta_{e2}-\frac{1}{2}\theta_{e1}+\theta_{t3} \nonumber \\
\delta d_{23}&=&\frac{2}{\sqrt{6}}\theta_s-\frac{1}{2\sqrt{3}}\theta_{e2}-\frac{1}{2}\theta_{e1}-\theta_{t3}  \nonumber
\label{normal2}
\end{eqnarray}

The spin-lattice interaction Hamiltonian becomes, then:

\begin{eqnarray}
H_{sl}&=&-\frac{2}{\sqrt{6}}g\theta_s(E_{12}+E_{34}+E_{13}+E_{24}+E_{14}+E_{23})  \nonumber \\
&-&\frac{1}{2}g\theta_{e1}(E_{13}+E_{24}-E_{14}-E_{23})   \nonumber \\
&-&\frac{1}{2\sqrt{3}}g\theta_{e2}(2E_{12}+2E_{34}-E_{13}-E_{24}-E_{14}-E_{23}) \nonumber \\
&-&g\theta_{t1}(E_{12}-E_{34})-g\theta_{t2}(E_{13}-E_{24})  \nonumber \\
&-&g\theta_{t3}(E_{14}-E_{23})   
\label{hspinlat}
\end{eqnarray}

\noindent where $E_{ij}$ denotes the energy of the $ij$ bond in $J$ units and $g\equiv \frac{\partial J(d)}{\partial d}|_{d=d_0}$, with $d_0$ the equilibrium distance.

In this way it is straightforward to have the energy gain related to each mode of our spin-orbital model. Treating separately all the cases of 
topologically distinct tetrahedra, we get:

{\bf 1)} $A$-tetrahedron with two spin singlets located at bonds $d_{12}$ and $d_{34}$. Superexchange energies are: $E_{12}=E_{34}=-4$ and
 $E_{13}=E_{24}=E_{14}=E_{23}=0$. The global magnetoelastic Hamiltonian $H_{meA}=H_{slA}+H_{elA}$ is: 

\begin{eqnarray}
H_{meA}=\frac{16}{\sqrt{6}}g\theta_{s}+\frac{1}{2}k_s\theta_{s}^2 
+\frac{8}{\sqrt{3}}g\theta_{e2}+\frac{1}{2}k_e\theta_{e2}^2
\label{meA}
\end{eqnarray}

As each normal mode is independent of the other, it is possible to minimize with respect to the 2 variables separately and get the energy minima:
 $E_{sm}=-\frac{64g^2}{3k_s}$, and $E_{e2m}=-\frac{32g^2}{3k_e}$. The normal modes corresponding to these minima are: 
$\theta_{sm}=-\frac{16g}{\sqrt{6}k_s}$, and $\theta_{e2m}=-\frac{8g}{\sqrt{3}k_e}$.
As $k_s=2k_t=4k_e=8k$, the global minimum is: $E_m=-8\frac{g^2}{k}$.

{\bf 2)} $B_1$-tetrahedron with superexchange energies: $E_{14}=E_{24}=-1$, $E_{13}=-4$ and $E_{23}=E_{12}=E_{34}=0$. 
The global magnetoelastic Hamiltonian is: 

\begin{eqnarray}
H_{meB2}&=&\frac{12}{\sqrt{6}}g\theta_{s}+\frac{1}{2}k_s\theta_{s}^2 
+2g\theta_{e1}+\frac{1}{2}k_e\theta_{e1}^2 \nonumber\\
&-&\sqrt{3}g\theta_{e2}+\frac{1}{2}k_e\theta_{e2}^2\label{meB2}\\
&+&3g\theta_{t2}+\frac{1}{2}k_t\theta_{t2}^2+g\theta_{t3}
+\frac{1}{2}k_t\theta_{t3}^2
\nonumber
\end{eqnarray}

The energy minima are: $E_{sm}=-12\frac{g^2}{k_s}$, $E_{e1m}=-2\frac{g^2}{k_e}$, $E_{e2m}=-\frac{3g^2}{2k_e}$, $E_{t2m}=-\frac{9g^2}{2k_t}$, 
and $E_{t3m}=-\frac{g^2}{2k_t}$. The normal modes corresponding to these minima are: $\theta_{sm}=-\frac{12g}{\sqrt{6}k_s}$, 
$\theta_{e1m}=-2\frac{g}{k_e}$, $\theta_{e2m}=\sqrt{3}\frac{g}{k_e}$, $\theta_{t2m}=-3\frac{g}{k_t}$, and $\theta_{t3m}=-\frac{g}{k_t}$.
The global minimum is: $E_m=-4.5\frac{g^2}{k}$.

{\bf 3)} $B_2$-tetrahedron with superexchange energies: $E_{14}=E_{12}=-1$, $E_{13}=-4$ and $E_{23}=E_{24}=E_{34}=0$. The global magnetoelastic Hamiltonian is: 

\begin{eqnarray}
H_{meB3}&=&\frac{12}{\sqrt{6}}g\theta_{s}+\frac{1}{2}k_s\theta_{s}^2 
+\frac{3}{2}g\theta_{e1}+\frac{1}{2}k_e\theta_{e1}^2 \nonumber \\ 
&-&\frac{\sqrt{3}}{2}g\theta_{e2}+
\frac{1}{2}k_e\theta_{e2}^2+ g\theta_{t1}+\frac{1}{2}k_t\theta_{t1}^2\nonumber \\ 
&+&4g\theta_{t2}+
\frac{1}{2}k_t\theta_{t2}^2+g\theta_{t3}+\frac{1}{2}k_t\theta_{t3}^2
\label{meB3}
\end{eqnarray}

The energy minima are: $E_{sm}=-12\frac{g^2}{k_s}$, $E_{e1m}=-\frac{9g^2}{8k_e}$, $E_{e2m}=-\frac{3g^2}{8k_e}$, $E_{t2m}=-8\frac{g^2}{k_t}$, 
and $E_{t1m}=E_{t3m}=-\frac{g^2}{2k_t}$. The normal modes corresponding to these minima are: $\theta_{sm}=-\frac{12g}{\sqrt{6}k_s}$, 
$\theta_{e1m}=-\frac{3g}{2k_e}$, $\theta_{e2m}=\frac{\sqrt{3}g}{2k_e}$, $\theta_{t2m}=-4\frac{g}{k_t}$, and $\theta_{t1m}=\theta_{t3m}=-\frac{g}{k_t}$.
The global minimum is again: $E_m=-4.5\frac{g^2}{k}$.

{\bf 4)} $B_3$-tetrahedron with superexchange energies: $E_{14}=E_{23}=-1$, $E_{13}=-4$ and $E_{24}=E_{12}=E_{34}=0$. The global magnetoelastic Hamiltonian is: 

\begin{eqnarray}
H_{meB1}&=&\frac{12}{\sqrt{6}}g\theta_{s}+\frac{1}{2}k_s\theta_{s}^2 
+g\theta_{e1}+\frac{1}{2}k_e\theta_{e1}^2 \nonumber \\
&-&\sqrt{3}g\theta_{e2}+\frac{1}{2}k_e\theta_{e2}^2+4g\theta_{t2}+\frac{1}{2}k_t\theta_{t2}^2
\label{meB1}
\end{eqnarray}

The energy minima are: $E_{sm}=-12\frac{g^2}{k_s}$, $E_{e1m}=-\frac{g^2}{2k_e}$, $E_{e2m}=-\frac{3g^2}{2k_e}$, and $E_{t2m}=-8\frac{g^2}{k_t}$. 
The normal modes corresponding to these minima are: $\theta_{sm}=-\frac{12g}{\sqrt{6}k_s}$, $\theta_{e1m}=-\frac{g}{k_e}$, 
$\theta_{e2m}=\sqrt{3}\frac{g}{k_e}$, and $\theta_{t2m}=-4\frac{g}{k_t}$.
The global minimum is still: $E_m=-4.5\frac{g^2}{k}$.

Due to the energy gain in the triplet $t_2$-mode, there is a shortening of the bond where the singlet is located, as 
it can be seen by substituting back the previous expressions for normal minima in Eqs. (\ref{normal})-(\ref{normal2}).

It is worthwhile to note that even though all three $B$ tetrahedra have the same magnetoelastic energy, the corresponding bond distortions are different, 
and thus the three configurations match one another differently when forced in the global fcc cell.

{\bf 5)} $C_1$-tetrahedron (the one shown in Fig. \ref{fig2}(e)) with superexchange energies: $E_{12}=E_{34}=0$ and $E_{13}=E_{24}=E_{14}=E_{23}=-1$. 
The global magnetoelastic Hamiltonian is: 

\begin{eqnarray}
H_{meC1}&=&\frac{8}{\sqrt{6}}g\theta_{s}+\frac{1}{2}k_s\theta_{s}^2\nonumber\\ 
&-&\frac{2}{\sqrt{3}}g\theta_{e2}+\frac{1}{2}k_e\theta_{e2}^2
\label{meC1}
\end{eqnarray}

The energy minima are: $E_{sm}=-\frac{16g^2}{3k_s}$, and $E_{e2m}=-\frac{2g^2}{3k_e}$. The normal modes corresponding to these minima are: 
$\theta_{sm}=-\frac{8g}{\sqrt{6}k_s}$, and $\theta_{e2m}=\frac{2g}{\sqrt{3}k_e}$.
The global minimum is: $E_m=-\frac{g^2}{k}$.

{\it 6)} $C_2$-tetrahedron (not shown in Fig. \ref{fig2}). They are characterized by 4 $b_1$-, one $b_2$- and one $b_3$-bonds, 
with superexchange energies: $E_{23}=E_{24}=0$ and $E_{13}=E_{14}=E_{12}=E_{34}=-1$. The global magnetoelastic Hamiltonian is: 

\begin{eqnarray}
H_{meC2}&=&\frac{8}{\sqrt{6}}g\theta_{s}+\frac{1}{2}k_s\theta_{s}^2 
+\frac{1}{\sqrt{3}}g\theta_{e2}+\frac{1}{2}k_e\theta_{e2}^2 \nonumber \\
&+&g\theta_{t2}+\frac{1}{2}k_t\theta_{t2}^2+g\theta_{t3}+\frac{1}{2}k_t\theta_{t3}^2
\label{meC2}
\end{eqnarray}

The energy minima are: $E_{sm}=-\frac{16g^2}{3k_s}$, $E_{e2m}=-\frac{g^2}{6k_e}$, and $E_{t2m}=E_{t3m}=-\frac{g^2}{2k_t}$.
 The normal modes corresponding to these minima are: $\theta_{sm}=-\frac{8g}{\sqrt{6}k_s}$, and $\theta_{e2m}=-\frac{g}{\sqrt{3}k_e}$,
 and $\theta_{t2m}=\theta_{t3m}=-\frac{g}{k_t}$.
As in the case of $B_i$ tetrahedra, the partial contributions of each mode compensate and the global minimum is, as for $C_1$ case, $E_m=-\frac{g^2}{k}$.



\begin{thebibliography}{99}
\bibitem{trig}
We neglect here the effects of the smaller trigonal field in the $t_{2g}$ manifold.
\bibitem{spinorbit}
If a complex linear combination is considered, there could be an associated spin-orbit energy gain. However, we have checked that the energy loss related to the breakdown of the spin-singlet state exceeds by far the possible spin-orbit energy gain.
\bibitem{nota}
In the case of ferro-orbital order there are in principle two $b_0$-bonds every tetrahedron. 
Yet, when the whole pyrochlore lattice is considered, only one of 
the two can be "active" as a spin-singlet.
\bibitem{note}
Due to the damping term $\Gamma$, the signal at a given energy involves an integral around that energy of width $\Gamma$ (see Eq. (\ref{eq_arxs})).
\bibitem{nota4}
This value is estimated from Ref. \onlinecite{natoli}.
\end{thebibliography}
\end{document}